\documentclass[secnumarabic,amssymb,nobibnotes,notitlepage, superscriptaddress,aps,prl,reprint, twocolumn]{revtex4-2}
\usepackage[english]{babel}
\usepackage[utf8]{inputenc}
\usepackage{amsmath,amssymb,bm}
\usepackage{ulem}
\usepackage{graphicx}
\usepackage{color}
\usepackage{comment}
\usepackage{eqnarray,amsmath}
\usepackage[colorinlistoftodos, color=green!40, prependcaption]{todonotes}
\usepackage[pdftex, pdftitle={Article}, pdfauthor={Author}]{hyperref} % For hyperlinks in the PDF
\usepackage[export]{adjustbox}
\usepackage[version=4]{mhchem}
\begin{document}
\title{Active polymers translocate faster in confinement}

%\author{K. R. Prathyusha, Paulami Sarkar, Justin Xu,   and M. Saad Bhamla} %(Order and contributions needs to be discussed)}

    %Harry Tuazon\email[Correspondence email address: ]{krprathyusha@gmail.com}% Your name
   % \affiliation{School of Chemical and Biomolecular Engineering, Georgia Institute of Technology}
\author{K. R. Prathyusha}
\email{krprathyusha@gmail.com}
\affiliation{ School of Chemical and Biomolecular Engineering, Georgia  Institute of Technology, Atlanta GA 30332, USA}
\author{ Paulami Sarkar}
\affiliation{ School of Civil and Environmental Engineering, Georgia Institute of Technology, Atlanta GA 30332, USA}
\author{Justin Xu}
\affiliation{School of Computer Science, Georgia Institute of Technology, Atlanta GA 30332, USA}
\author{Saad Bhamla}
\email{saadb@chbe.gatech.edu}
\affiliation{ School of Chemical and Biomolecular Engineering, Georgia Institute of Technology, Atlanta GA 30332, USA}

%\email{ramin.golestanian@ds.mpg.de}

%\date{\today}
%\date{\today} % Leave empty to omit a date

\begin{abstract}
Living organisms employ diverse strategies to navigate confined environments. Inspired by translocation observations on California blackworms (\textit{Lumbriculus variegatus}), we combine biological experiments and active-polymer simulations to examine how confinement and stiffness govern translocation. Active filaments translocate fastest when the channel width is comparable to their diameter, with escape time determined by propulsion speed, filament length, and channel geometry. In wider channels, activity and flexibility induce reorientation-dominated conformational changes that prolong escape. A single dimensionless ratio linking confinement to stiffness captures the transition from axis-aligned escape with short wall deflections for stiffer filaments, to  reorientation-controlled motion with blob-like shapes for flexible filaments. These results provide a unified physical framework for active translocation in confinement and suggest design principles for flexible robotic filaments in complex environments.
\end{abstract}
\maketitle
%\section{introduction}
%\section{Abstract}
%Translocation through confined channels demonstrates non-equilibrium behaviours in active filaments and externally driven passive filaments. 
%Active filaments and  passive filaments exhibit  translocation through confined channels.
%Translocation of passive flexible filaments, such as DNA, RNA, and proteins, through geometries like nanopores, nuclear pores, or microchannels is inherently a non-equilibrium process driven by external forces and constraints, and it involves conformational changes influenced by their flexibility, length, and interactions with confining walls
%Active and passive filaments undergo translocation through confined channels, a process involving the navigation of filaments through  confined geometries, which is fundamental to various phenomena across scales 
%Active and passive filaments undergo translocation through confined channels, a process which is fundamental to various phenomena across scales 
\textit{Introduction---\;}Translocation of passive and active flexible filaments through confined channels is fundamental  across scales \cite{muthukumar-book-PT,panja-review-2013,das-blobscaling-prl2019}. Owing to their thermal nature, passive filaments (e.g. DNA, RNA, proteins)  typically require external forces or gradients to traverse  nanopores and cellular channels. 
Their dynamics are governed by chain flexibility ~\cite{sung-pt-throughpore-PRL-96,muthukumarPRL-translocation-2001,muthukumar-book-PT}, polymer-pore interactions ~\cite{polymer-pore-luo-prl-07,chen-attr-pore-polymer-jcp-09},  confinement geometry ~\cite{cacciuto-confinement-prl-06,mohan-jcp-polymer-10,pt-geometry-pre-19}, and external driving \cite{luo-pt-efield-jcp-06}, relevant to DNA sequencing ~\cite{dna-confinement-review-2012}, gene delivery, and nanotechnology~\cite{ying2022nanopore}.

In contrast to passive polymers, active filaments  %and snakes~\cite{rieser2021functional-snake}  
consume energy to sustain persistent movement and undergo shape fluctuations driven by internal forces. The coupling of activity and flexibility gives rise to distinctive conformational dynamics~\cite{fjnedelec-97,Pfreundt2023,wormreview,prathyusha_softmatter-22,riseleholder-15,lowenpolymer-swellingactivebath2014,akaiser-15} and collective behaviors~\cite{prathyusha2018PRE,llgof-02,fjnedelec-97,leiladirectionreversal2023,harryscience2023}. 
Examples span  cytoskeletal filaments in cells and macroscopic organisms, including earthworms  and snakes navigating pores and crevices.
Recent work on transport of active agents in dense or confined media has focused on active colloids~\cite{Bechinger2016, prathyusha2024anomalous}  and rod-like bacteria~\cite{suvendunat-com}. By contrast, the translocation of  flexible active filaments through confinement remains comparatively underexplored~\cite{active-filament-in-crowded-media-23}.

Passive polymers in confinement are well described by Odijk’s deflection-segment picture for narrow channels~\cite{odijk1983statistics}
and de Gennes's blob model for wider ones~\cite{degennes-blob-77}. A recent study shows that such classical scalings can break down for active filaments when the activity-induced persistence length approaches the blob size~\cite{das-blobscaling-prl2019}. Furthermore, simulations and experiments  on active polymer translocation  show that activity speeds up transport~\cite{tan-length-width-2d-2023} and modifies how escape time scales with polymer size~\cite{hu2024translocation,Antoine-translocation-25}, stiffness~\cite{tan-forced-2024}, and conformation~\cite{locatteli-filament-channel}, producing behaviors absent in confined passive systems. Because propulsion mechanisms differ across  self-propelled filaments, no single scaling applies. For snake-like locomotion, anisotropic friction with the substrate can slow motion as the channel narrows~\cite{snake-channel-Dhu2012}; in contrast, worms generate motion primarily through internal actuation with limited wall friction~\cite{wormreview}, so their confined transport warrants  separate examination~\cite{biswas2023escape,Antoine-translocation-25}. This motivates the central question we address here: how do channel width, flexibility, and activity  together govern the  active-filament transport?  Beyond its   fundamental interest for confined active matter, this question is relevant to 
 the design of soft robotic filaments for  pipe inspection~\cite{ikeuchi2012development}, navigation in tortuous tubular environments~\cite{tang2022pipeline-robot}, and subsurface exploration~\cite{das2023earthworm,liu2023actuation-review}.
\begin{figure*}    \includegraphics[width=1\textwidth]{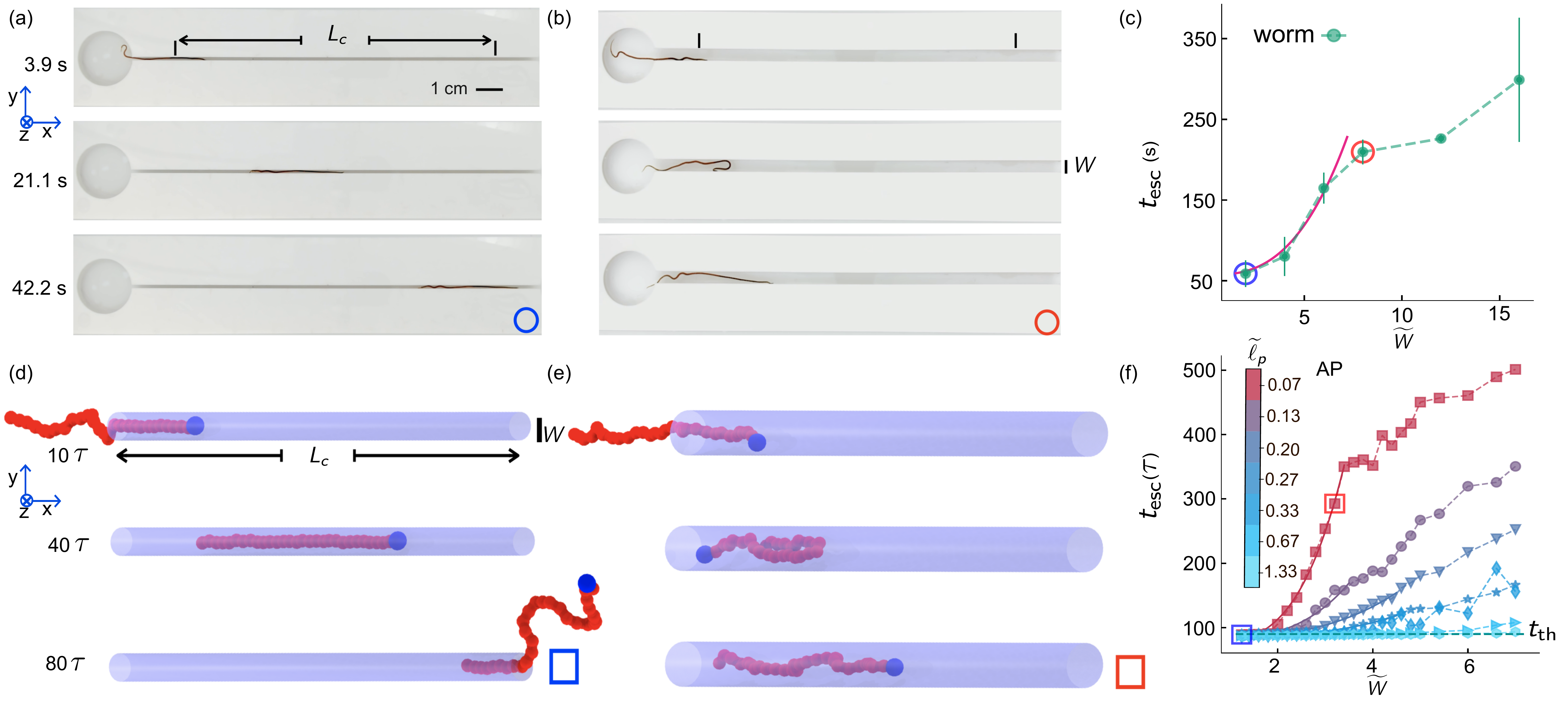}%
%\vspace{-15pt}
    \caption{{\bf Translocation  and escape time  of active filaments in confined channels}:   
 (a-b) Experiments: at scaled width ($\widetilde{W}=2$), a blackworm escapes within 
 $t_{esc}=56\pm16$ seconds; in a wider channel ($\widetilde{W}=8$),
 it traverses only half  the channel over the same  interval. (d-e) Simulation snapshots for active polymer translocation ${\widetilde{\ell_p}=0.07}$: at $\widetilde{W}=1.4$, strong confinement aligns the filament and speeds escape; in wider channels $\widetilde{W}=3.0$, frequent reorientations and lateral exploration slow transport. (c,f) Escape time measurements from both biological experiments and AP simulations showing that the  escape time scales with  the effective channel width, $t_{esc}= \widetilde{W}^\alpha+C$:  worms (solid line) $\alpha=1.8$; AP simulations $\alpha=[2.8-3.4]$ with $C \sim t_{th}$   for $\widetilde{\ell_p}\le 0.5$.  In AP simulations, at higher stiffness, $t_{esc}$ becomes nearly width-independent and approaches $t_{th}$. }
 %However, at larger bending rigidity $\kappa$, the escape dynamics remain largely unchanged with respect to channel width , indicating a deviation from the scaling behavior in the stiff polymer limit. }
    \label{fig1:translocation}
\end{figure*}
In this letter,  we investigate how channel width controls the translocation of flexible active  filaments using California blackworms (\textit{Lumbriculus variegatus}) experiments and Brownian-dynamics simulations of a  tangentially propelled polymer model. \\
\textit{Experimental system---\;}California blackworms are aquatic, slender organisms~(diameter $D=0.5\pm0.1~\textrm{mm}$, length $L=50\pm10~\textrm{mm} $) that are both flexible and motile,  making them a convenient model  for  active filaments in  open-ended channels (Fig.~\ref{fig1:translocation}, SI.~I-II)~\cite{wormreview}. The channel connects two reservoirs and is open at both ends; worms can enter and exit from either side, eliminating directional bias. 
%The experimental setup, comprising of  a transparent horizontal channel with two open ends made of acrylic material and having a length of L=16 cm,  height 1=mm, is filled with water (Figure \ref{fig:fig1setup}(a)). To minimise capillary effects, the top portion of the channel has a slit-like opening, ensuring that the dynamics observed are not influenced by surface tension. 
%The worms can escape and enter from both ends of the channel, ensuring no  directional bias in their movement. By systematically varying the channel width $W$ (THE VALUES), we can investigate the influence of spatial confinement on their behavior, facilitating direct observations of translocation \\
\\
\textit{ Active polymer (AP) simulation---\;}We simulate an active polymer composed of N monomers (diameter $\sigma$) with coordinates ${\bf r}_i$, propelling along the local tangent, connected by harmonic springs and  bending potential,  confined within a cylindrical channel. Each monomer $i$ obeys  overdamped dynamics, 
%\begin{equation}
$\gamma{{\dot {\bf r}}_i}=-\nabla_i({ U})+{\bf F}_i^{R}+ {\bf F}_i^{A}
 \label{eq:eqofmotion}$, where $\gamma$ is the damping constant and $U$ includes  stretching, bending, excluded-volume, and  wall-repulsion potentials~(SI. III). ${\bf F}_i^{R}$ is the stochastic force with zero mean and variance $2\gamma k_BT$; and the active force on each bead $i$  is ${\bf F}_i^{A}$ with magnitude $f_p$ directed along the tangent, i.e., 
$\mathbf{F}_i^{A}=\tfrac{f_p}{2}\left(\hat{\mathbf{t}}_{i-1,i}+\hat{\mathbf{t}}_{i,i+1}\right)$ for $i=2,\dots,N-1$, 
and $\mathbf{F}_1^{A}=f_p\hat{\mathbf{t}}_1,\ \mathbf{F}_N^{A}=f_p\hat{\mathbf{t}}_{N-1}$ for the chain ends. This minimal model serves as a controlled counterpart to the worm experiments by capturing key worm dynamics~\cite{sweeper-worm-collection-25,Antoine-translocation-25,nguyen2021emergent} and facilitating simulations of self-propelling filaments with tunable motility and flexibility\cite{prathyusha2018PRE,prathyusha_softmatter-22,prathyushatransvercargo,Paolo-2018globulelike}.

We systematically vary the channel width $W$ to probe how confinement affects translocation~(Fig~\ref{fig1:translocation}). The channel length is $L_c=12$ \textrm{cm} in experiments and $L_c=60 \sigma$ in AP simulations. In AP simulations, we choose 
 $\varepsilon$ and $\sigma$ as  energy and length units, respectively, and  define the time unit $\tau=\sigma^2\gamma/\varepsilon$ (SI. III).
Two dimensionless parameters  characterize the problem: the scaled  width $\widetilde
{W}=W/D$,  where $D$ is the filament diameter, indicating available lateral space, and the dimensionless stiffness  $\widetilde{\ell_p }=\ell_p /L$, which characterizes the chain flexibility.  In simulations, $\ell_p=\frac{\kappa}{k_BT}$ is the thermal persistence length, where $\kappa$ denotes the bending rigidity;  in experiments, $\ell_p$  is estimated from tangent–tangent correlations~\cite{sweeper-worm-collection-25}. We also use the P\'eclet number  $\mathrm{Pe} = \dfrac{f_{p}\,\sigma}{k_B T}$,  set to $10$. We fix $N=30$ and $k_BT=0.1\varepsilon$.
 \textit{Results---\;}We define a successful translocation as an event in which a filament  enters the channel from one end and fully exits the opposite end within the  observation window. The escape time $t_{\mathrm esc}$ is measured from head entry to tail exit, capturing the time required for the entire contour to traverse the channel~\cite{yong2012driven} (SI. IV, \cite{SM}).
 %\textit{Results---} In our experiments and simulations, a filament is said to undergo translocation when it enters the channel from one side and exits through the other end of the channel. The time required for this process, known as the translocation time or escape time ($t_{\mathrm esc}$), is measured as the duration between the initial entry of the head of the filament into one end of the channel and the complete exit of its tail from the opposite end. This method naturally captures the time required for the entire contour length to pass through the channel~\cite{yong2012driven}.
   \begin{figure*}
    \centering
    \includegraphics[width=1\linewidth]{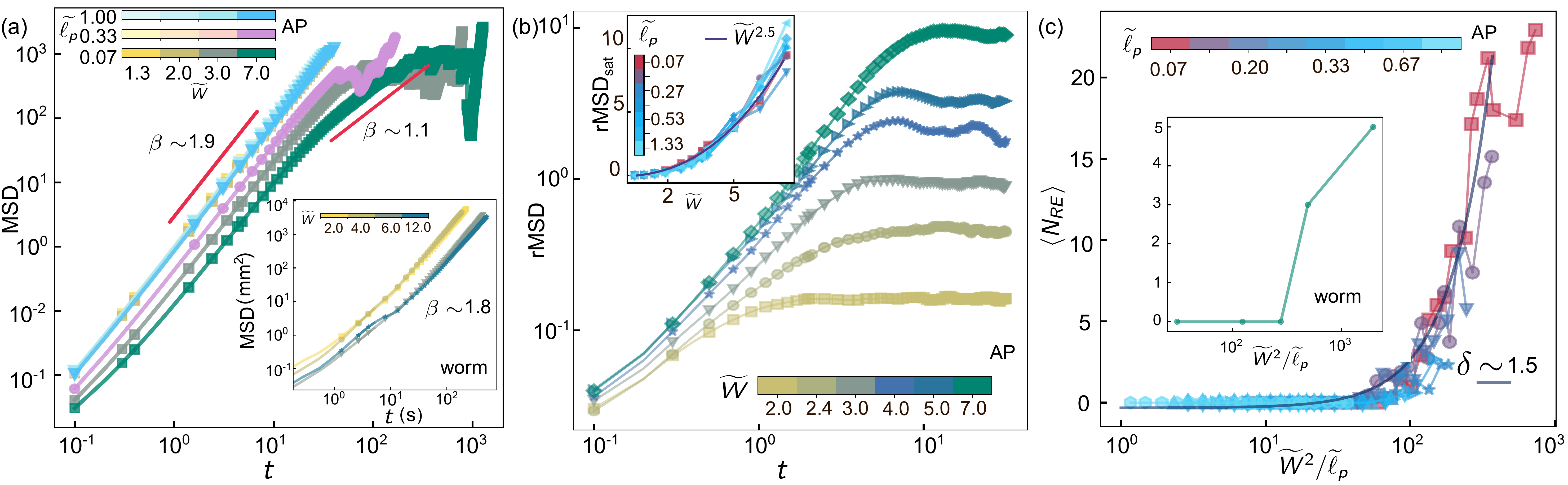}%
\caption{{\textbf{A single parameter ($\widetilde{W}^2/\widetilde{\ell}_p$) controls active translocation.}  (a) Mean-squared displacement (MSD) of the center of mass versus time for varying $\widetilde{W}$ and $\widetilde{\ell_p}$. Decreasing  $\widetilde{W}$ increases MSD, indicating enhanced transport  in simulations  and experiments (inset).
%  The exponent $\alpha$ takes the values $\alpha \approx 2.0$ ($\widetilde{W} < 2.0$) and $1.5 < \alpha < 2.0$ when $\widetilde{W} \ge 2.0$. 
(b) Transverse or radial rMSD, shown for various confinement widths for $\widetilde{\ell_p}=0.07$, grows then saturates; the saturation level (shown in inset, $\textrm{rMSD}_\textrm{sat}$) increases with $\widetilde{W}$ and is largely stiffness-independent, showing that radial exploration is not rate-limiting. (c) The average axial reorientation count $\langle N_{\mathrm{RE}}\rangle$, obtained from sign changes of the end-to-end vector $\mathbf{R}_E$, collapses onto a single curve when plotted against $\widetilde{W}^2/\widetilde{\ell}_p$, identifying a single control parameter that links lateral space to stiffness and rationalizes the escape slowdown in wide channels; experiments show similar trends (inset).
 %Mean squared displacement (MSD) of the polymer's center of mass as a function of time for different confinement strengths. Under strong confinement, the polymer exhibits ballistic-like motion with an exponent $\alpha \approx 2.0$, while under weak confinement, increased lateral fluctuations lead to super-diffusive behavior  with $1.5 < \alpha 2.0$. The inset shows the scaling exponent $\alpha$ for various $\kappa$ illustrating the transition from ballistic to super-diffusive motion with decreasing confinement strength.
 }}
    \label{fig:fig3msd}
\end{figure*}

Fig.~\ref{fig1:translocation}(a)  shows that in narrow channels ($\widetilde{W} <= 2$; strong confinement)  worms align with the channel axis and escape within $t_{\mathrm esc}=56\pm16$ seconds, with minimal transverse shape fluctuations.  In  wider channels ($\widetilde{W} \ge 2$), they  adopt undulatory shapes  and undergo frequent reorientations, yielding $t_{\mathrm esc} = 299\pm 77$ seconds (5x longer, Fig.~\ref{fig1:translocation}b). 
%When the channel is narrower ($\widetilde{W} < 2  $) (strong confinement),  (1 mm) wide, the filament escapes in  0.56 [UPDATE] minutes. In contrast, in a wider ($\widetilde{W}  \ge 2 $) (weak confinement) 6 mm channel, the filament only makes it halfway through the channel within the same time frame. 
%This stark difference highlights the  role of confinement:
These observations indicate stronger confinement enhances directed translocation and shortens escape time. % Thigmotaxis, Moreover,  the positive thigmotaxis of the worm, the tendency of the worm to stay in contact with surfaces, helps to maintain this alignment %along the channel walls, 
%and 
%reduces random re-orientations while enhancing forward propulsion. In wider channels ($\widetilde{W}>2.0$, Fig.~\ref{fig1:translocation}(b)),  the worm exhibits increased conformational fluctuations and re-orientations that disrupt persistent movement and slow down escape.
%This alignment is driven by thigmotaxis—the tendency of worms to stay in contact with surfaces—which plays a crucial role in guiding their movement along the channel walls. Blackworms exhibit positive thigmotaxis, preferring to remain in contact with the channel boundaries. However, despite this tendency, they still successfully escape, suggesting that confinement enhances alignment while allowing for directed propulsion. By maintaining contact with the confining boundaries, worms navigate more effectively, reducing random reorientations and enhancing directed motion. In wider channels, where the worm has more room to explore, it experiences increased conformational fluctuations and undergoes multiple reorientations due to self-propulsion and wall interactions, leading to frequent undulations and directional changes that slow down the translocation process. 

AP simulations reproduce these behaviors (Fig.~\ref{fig1:translocation}d,e). In narrow channels, $\widetilde {W} \le 2 $,  steric interactions align the active filament with the axis, yielding  directed motion and faster escape. For 
wider channels, $\widetilde {W} > 2 $, frequent reorientations and conformational fluctuations slow transport,  consistent with experiments. In this regime, flexible filaments exhibit blob-like conformations reminiscent of de Gennes–like picture~\cite{degennes-blob-77} and  reorientation-controlled dynamics. Whereas stiffer filaments form short wall-deflection segments and remain axis-aligned, reminiscent of an Odijk-like picture~\cite{odijk1983statistics}, which reduces reorientations and favors   directed escape. 

%We measure the translocation time or escape time ($t_{\mathrm {esc}}$) as the interval  between the initial entry of the filament head and the exit of the filament tail from the opposite end, naturally capturing the time required for the entire contour to traverse the channel ~\cite{yong2012driven} (SI. IV). 
%Fig.~\ref{fig1:translocation}(c)\&~\ref{fig1:translocation}(f) present the escape time measurements from both experiments and simulations, revealing a clear confinement-dependent trend. 
As shown in Fig.~\ref{fig1:translocation}c, worm translocation time increases with the scaled channel width $\widetilde{W}$; worms therefore  escape more quickly in narrower channels. In simulations, for ($\widetilde{W} \le 2$), confinement suppresses transverse fluctuations 
and the filament adopts a nearly straight conformation, effectively moving as a rigid rod at speed $f_p/\gamma$.
In this strong-confinement limit, $t_{\mathrm esc}$ becomes insensitive to flexibility and thermal noise and approaches $t_{\mathrm{th}} = \gamma(L_c + L)/f_p$, where $L = (N - 1)\sigma$; accordingly, $ t_{\mathrm{esc}}$ is nearly width-independent and close to $t_{\mathrm{th}}$  (Fig.~\ref{fig1:translocation}f).

%Assuming a constant propulsive force and negligible thermal fluctuations, the filament moves axially, and the measured

For wider channels ($ \widetilde{W} > 2 $), both worms and simulations exhibit longer escape times due to increased exploration with frequent reorientations and shape changes (Fig.~\ref{fig1:translocation}c,f).  %as observed in both experiments and simulations.
  %When ($ \widetilde{W} > 2 $) the filament has more space to undergo transverse fluctuations, which promote reorientation driven by self-propulsion and result in an increase in translocation time with channel width, as observed in both experiments and simulations. 
  In both cases, the data are described by a power-law scaling $t_{\mathrm{esc}} = \widetilde{W}^\alpha+ C $; for worms $\alpha \sim ~1.8$, and in simulations $\alpha\approx 2.8-3.4$ with $C\approx t_{th}$, reflecting the rigid, propulsion-limited offset. Notably, these behaviors differ from passive‑polymer translocation, where confinement adds entropic resistance without significant reorientation‑controlled delays~\cite{muthukumar-book-PT}, and they also contrast prior studies on active filaments in confinement~\cite{tan-length-width-2d-2023}, either in 2D geometries or under friction-dominated motion~\cite{rieser2021functional-snake}.

To characterize  filament motion, we measure the mean-squared displacement of the center of mass (${\bf r}_{cm}$) of the polymer, $ \mbox{MSD}(t)= \langle |{\bf r}_{cm}(t)-{\bf r}_{cm}(0)|^2 \rangle$ $\sim t^{\beta}$. Here, $ \langle\cdot\rangle$ denotes averaging over  intervals when the filament is fully inside the channel and over successful translocations.
%Where $t = 0$ marks the time the center of mass enters the channel and   $\langle.. \rangle$ denotes  averaging over successful translocations. %Where the time origin corresponds to the moment when the center of mass of the polymer is inside the channel, and averaging is performed only over successful translocation runs. 
%The MSD scaling provides insight into the nature of the polymer motion, characterized by MSD $\sim t^\alpha$.
Decreasing $\widetilde{W} $ increases the MSD amplitude, indicating
enhanced transport for both the AP model and worms (Fig~\ref{fig:fig3msd}a, inset for worms). For ($\widetilde{W} \le 2$),  motion  is predominantly deterministic rather than diffusive, with  super-diffusive scaling $\beta\sim 1.9$ that is insensitive to stiffness $\widetilde{\ell}_p$. In wider channels, flexible filaments  crossover from  superdiffusive to near-diffusive dynamics ($\beta\sim 1.1$) as reorientations  decorrelate directional persistence, and the MSD saturates at long times due to confinement;  stiff filaments remain superdiffusive, consistent with axis-aligned Odijk-like conformations and directed escape. For worms, the dynamics stays superdiffusive while the MSD amplitude decreases in wider channels, in qualitative agreement with the AP model (Fig~\ref{fig:fig3msd}a inset). A short intermediate subdiffusive plateau appears in worms due to intermittent pauses of the worm, a feature absent in the AP model because propulsion is constant (SI V).

 We next quantify transverse exploration by the radial MSD of  monomers,
 rMSD $= \langle \frac{1}{N} \sum_{i=1}^{N} 
| \mathbf{r}_i^{\perp}(t+\tau) - \mathbf{r}_i^{\perp}(\tau) |^2 \rangle$, with $\mathbf{r}_i^{\perp}(t) = \{ y_i(t), z_i(t) \}$ (Fig.~\ref{fig:fig3msd}b). The rMSD grows and then saturates at a level $\textrm {rMSD}_{\textrm{sat}}$ set by the channel width. %$\Delta {\mathbf r}^2_{\perp, {\mathrm sat}}$ grows systematically with channel width due to activity-driven exploration of the confinement boundary and remains largely independent of filament flexibility. 
 $\textrm {rMSD}_{\textrm{sat}}$ increases with $\widetilde{W}$  and  is largely independent of stiffness $\widetilde{\ell_p }$ (inset Fig.~\ref{fig:fig3msd}b), with $\textrm{rMSD}_{\textrm{sat}} \sim \widetilde{W}^{2.5}$,  indicating that self-propulsion enhances transverse fluctuations beyond passive geometric expectations.

Since the saturation time for radial MSD is independent of flexibility (SI VI, Fig S4), delays in translocation are likely set by  axial reorientations. We quantify these reorientations from the end-to-end vector ${\bf R}_{E}(t)$=${\bf R}_N(t)-{\bf R}_1(t)$, via its alignment with the channel axis, ${sgn}{[\bf \hat R}_E(t) \cdot \hat{\bf x}]$.
%We quantify those reorientations by the time evolution of the end-to-end vector of the polymer, ${\bf R}_{E}(t)$=${\bf R}_N(t)-{\bf R}_1(t)$ and  its alignment with the channel axis  $\hat{\bf x}$, i.e., ${sgn}[{\bf \hat R}_E(t) \cdot \hat{\bf x}]$.
%. We measured how often the filament reverses its axial alignment by computing the sign of the dot product between ${\bf R}_{E}(t)$ and the unit vector  of the channel axis~$\hat{\bf x}$, i.e., ${sgn}[{\bf R}_E(t) \cdot \hat{\bf x}]$. 
Each sign change marks an axis reversal, and the total count over a trajectory gives $N_{RE}$.

%of $\bf{R}_E$ %relative to $\hat{\mathbf{x}}$,
%and counting such sign changes over a translocation trajectory yields the total number of axial reversals.
%the total number of sign changes over the translocation trajectory provides the  number of the  axial reversals. %of the filament.

To rationalize the dependence on confinement and stiffness,  we introduce the dimensionless control parameter, $\widetilde{W}^2/\widetilde{\ell_p}$, which captures the competition between lateral exploration ($\sim\widetilde{W}^2$) and orientational persistence (set by $\widetilde{\ell_p}$). 
This scaling  arises from the fact that the polymer has roughly $L/\ell_p$ independent segments, and the reorientation probability in each segment increases with the available lateral space $\widetilde{W}^2$, leading to the control parameter $ \widetilde{W}^2/\widetilde{\ell_p}$.
The average axial reorientation count per successful translocation, $\langle N_{\mathrm{RE}}\rangle$,   collapses when plotted against  $\widetilde{W}^2/\widetilde{\ell_p}$ (Fig.~\ref{fig:fig3msd}c), indicating a single-parameter organization of the dynamics across widths and stiffnesses.
%determine the likelihood of reorientation during translocation.
%Wider channels increase the geometrically accessible fluctuation range ($ \sim \widetilde{W}^2$), while the ability of the  filament to maintain its orientation is set by $\widetilde{\ell_p}$.
%he average re-orientation number $\langle N_{{RE}} \rangle$ against the  parameter, $\widetilde {W}^2/\widetilde{\ell_p}$, exhibiting a  universal collapse of data for different flexibility and confinement onto a single curve. The observed collapse occurs because  $\widetilde {W}^2/\widetilde{\ell_p}$  captures the balance between lateral confinement and filament flexibility, which together determine the likelihood of reorientation during translocation.  

For  $ \widetilde{W}^2/\widetilde{\ell_p}< 10^2$, $\langle N_{{RE}} \rangle\approx 0$,  consistent with strong confinement ($\widetilde{W}\le2$) or an Odijk-like, axis-aligned regime. As $ \widetilde{W}^2/\widetilde{\ell_p}$ increases beyond the threshold, $\langle N_{{RE}} \rangle$ rises, signaling a transition to a de Gennes-like regime, where increased lateral space and flexibility promote frequent reorientations and longer residence times. In this regime, $\langle N_{{RE}} \rangle \sim (\widetilde{W}^2/\widetilde\ell_p)^\delta$ with $ \delta=[1.5 - 1.7]$, corresponding to $\langle N_{{RE}} \rangle \sim \widetilde{W}^{[3.0{-}3.4]}$  at fixed $\widetilde{\ell_p}$. 
Although the escape time shows a similar growth ($t_{\mathrm{esc}} \sim \widetilde{W}^{[2.8{-}3.4]}$), many reorientations are small-angle events that do not fully interrupt forward motion.
%Despite  this increase in $\langle N_{\mathrm{RE}}\rangle$,   %reorientations ($\langle N_{\mathrm{RE}} \rangle \sim W^{3.0{-}3.4}$),
%translocation time grows more moderately as $\tau_{\mathrm{esc}} \sim W^{2.8{-}3.2}$, 
 %reorientations involve small angular deviations  that do not  hinder forward motion.
 Nevertheless, the observed scaling of $t_{\mathrm{esc}}$ and $\langle N_{{RE}} \rangle$ indicates that reorientation is the primary source of slowdown for flexible filaments in wider channels.
%suggests that reorientations are the dominant contributor to the slowdown of  flexible filaments in wider channels.
Worms exhibit many reorientations overall but fewer during successful translocations; despite limited statistics,  they show the same qualitative trend (Fig.~\ref{fig:fig3msd}c, inset).

\begin{figure}
    \centering
\includegraphics[width=0.5\textwidth]
{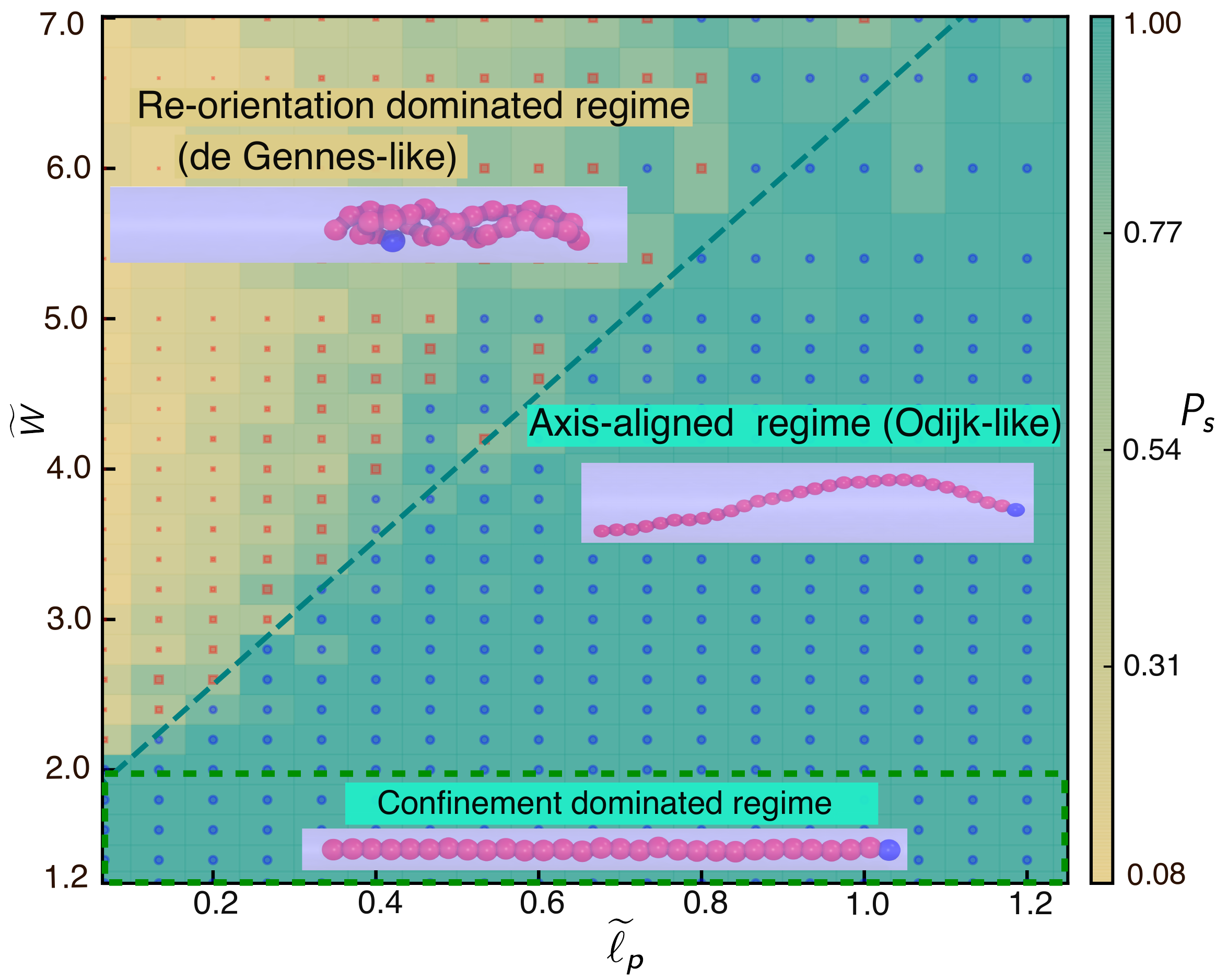}%
%\vspace{-15pt}
\caption{{ \textbf{A simple confinement–stiffness boundary predicts escape success.} Translocation success probability $P_s$  versus scaled width $\widetilde{W}$ and stiffness $\widetilde{\ell}_p$. Each symbol represents $200$ independent runs, and its area scales with  $P_s$. The dashed  line  ($P_s = 0.85$) gives an empirical boundary, $\widetilde{W} = 4.83\widetilde{\ell}_p + 1.6$, separating high-success, axis-aligned, Odijk-like transport from low-success, reorientation-dominated, de Gennes-like motion  for $\widetilde{W} >2$. Strong confinement ($\widetilde{W} \le 2$}) yields $P_s\approx1$. Representative conformations illustrate both regimes.}
    \label{fig:fig4ps}
\end{figure}

In addition to slowing  escape, the interplay of filament flexibility and confinement also affects the likelihood of successful translocation.
We define the success probability $P_s = N_s / N_t$, where $N_s$ is the number of successful exits  and $N_t$  the  number of  entries. Failed events ($N_t-N_s$) include exits from the entry side or trapping by excessive reorientations, entanglements, or back-and-forth motion during the observation time.
%becomes trapped due to excessive reorientations, entanglements, or back-and-forth motion that halts forward motion.
 %In weak confinement, the filament slows down not only due to the increased available space but also because of frequent reorientations and tumbling behaviour. These dynamics reduce the number of successful translocations, particularly for flexible filaments. To quantify this, we define the  success probability as~${ P}_{s}=\frac{N_s}{N_t} $. Here, $N_s $ is the total number of successful trials, where the filament fully exits from the channel and $N_t$ is the total number of  trials  where the filament enters the channel.   Failed events include cases where the filament either exits from the entry side or becomes trapped due  to excessive reorientations, conformational entanglements, or back-and-forth motion, preventing forward progress over the entire simulation period or experimental observation time. 
 As shown in Fig.~\ref{fig:fig4ps}, the success probability
  is high ($P_s=1$)  for $\widetilde{W}\le 2$, independent of filament stiffness $\widetilde{\ell_p }$. %, as geometric constraints suppress fluctuations and prevents backward motion.
  % across all filament flexibilities. In this regime, steric constraints dominate, effectively suppressing transverse fluctuations and preventing backward motion, regardless of filament stiffness.
  However, for $\widetilde{W}>2.0$, translocation success depends strongly on filament flexibility: Odijk-like, axis-aligned conformations in stiffer filaments yield high success rates ($  P_s \gtrsim 0.85$),  whereas, flexible filaments in wider channels enter a de Gennes–like regime dominated by reorientations and shape fluctuations, reducing $P_s$.
The dashed boundary $P_s=0.85$ (green line) separates the Odijk-like and de Gennes–like regimes and summarizes the design space for successful translocation.
 
% The green line ($\widetilde{W} = 4.83\widetilde{\ell}_p + 1.6$) separates the  Odijk-like regime from the  de-Gennes like regime, highlighting a key feature of active polymer dynamics under confinement.

\textit{Outlook---\;}Flexibility, activity, and confinement jointly determine active-filament translocation. Using  worms and Brownian-dynamics simulations, we show that narrow channels promote axis alignment and rapid escape, while wider channels permit   reorientations and conformational changes that  hinder transport. Unlike passive polymer translocation governed by diffusion and entropic forces, active filaments 
exhibit confinement- and flexibility-controlled motion that bridges Odijk-like, axis-aligned transport and de Gennes–like reorientation-dominated escape. 

 The minimal tangentially propelled-filament model used here captures the key ingredients  of active  translocation and  highlights how dimensionality modifies confinement-driven dynamics. Future work should incorporate hydrodynamics,  boundary compliance and friction, branched/curved geometries, and multi-filament interactions to test the generality of the mechanism and the single-parameter collapse. More broadly, these results motivate  studies of navigation strategies in flexible organisms, cytoskeletal filaments, and synthetic active matter in complex environments. Finally, the success-probability map and the confinement-stiffness boundary provide physical design rules for autonomous soft robots and microrobot chains  navigating confined,  irregular pathways~\cite{veenstra2025adaptive,huang2019adaptive}, with applications in pipe inspection~\cite{pipereview.2023.49}, drug delivery~\cite{li2017micro}, and medical diagnostics~\cite{ren2021softSciAdv}.
\section{acknowledgments}
We  thank Saeed Mahdisoltani, Harry Tuazon, Arnold Mathijssen, Aparna Bhaskaran, Daehyun Choi and Kang Sup Lee  for helpful discussions. We acknowledge the resources and services provided by the Partnership for an Advanced Computing Environment (PACE) at the Georgia Institute of Technology, USA. S.B. acknowledges support from NSF awards PHY-2310691 and iOS-1941933; National Institutes of Health (NIH) award R35GM142588 and Schmidt Sciences, LLC.

\newpage
\widetext
\makeatletter
\renewcommand{\fnum@figure}{\figurename~S\thefigure}
\makeatother

\begin{center}
\textbf{\large Supplemental Materials:~ Active polymers translocate faster in confinement}
\end{center}
\setcounter{figure}{0}

\renewcommand{\theequation}{S\arabic{equation}}

%\title{Enhanced  translocation of active filaments in confinement}
%\title{Supplemental material for Active polymers translocate faster in confinement}
%\renewcommand{\thefigure}{S\arabic{figure}} % makes numbers S1, S2, ...

%\author{K. R. Prathyusha, Paulami Sarkar, Justin Xu,   and M. Saad Bhamla} %(Order and contributions needs to be discussed)}

    %Harry Tuazon\email[Correspondence email address: ]{krprathyusha@gmail.com}% Your name
   % \affiliation{School of Chemical and Biomolecular Engineering, Georgia Institute of Technology}
%\author{K. R. Prathyusha}
%\email{krprathyusha@gmail.com}
%\affiliation{ School of Chemical and Biomolecular Engineering, Georgia Institute of Technology, Atlanta GA 30332, USA}%
%\author{ Paulami Sarkar}
% \affiliation{School of Civil and Environmental Engineering, Georgia Institute of Technology, Atlanta GA 30332, USA}
%\author{Justin Xu}
%\affiliation{School of Computer Science, Georgia Institute of Technology, Atlanta GA 30332, USA}
%\author{ M. Saad Bhamla}
%\email{saadb@chbe.gatech.edu}
%\affiliation{ School of Chemical and Biomolecular Engineering, Georgia Institute of Technology, Atlanta GA 30332, USA}
%%\email{ramin
%\newcommand{\KRP}[1]{\textcolor{black}{#1}}

%\date{\today}

%\pacs{Valid PACS appear here}% PACS, the Physics and Astronomy

%\maketitle
%\tableofcontents
%\newpage
In this supplement, we detail  the experimental setup, image-analysis procedures, AP simulation model and parameters, translocation-time definitions, and additional analyses (MSD exponent, radial‑MSD saturation, and reorientation metrics). We also provide representative movies from  experiments and simulations that support the main results.

\section{Experimental Setup}

\begin{figure*}[!h]
    \centering
\includegraphics[width=\textwidth]{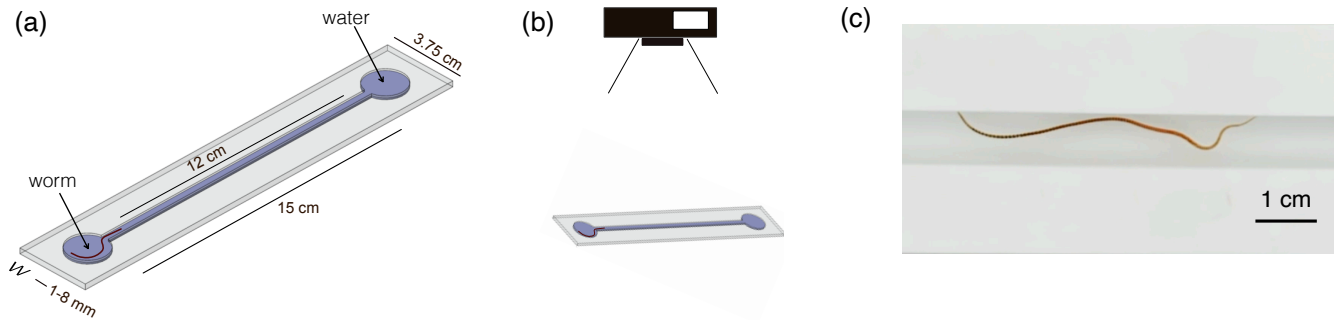}
    \caption{ 
   \textbf{Schematic of experimental setup.} (a) Quasi-2D setup used for worm translocation experiments:  two circular chambers connected by an open-ended, narrow channel (dimensions annotated in schematic). (b) Imaging setup: a top-mounted camera records motion of the  worm  between  chambers under fixed illumination. (c) Image of California blackworm (\textit{Lumbriculus variegatus}) inside the channel.    
   }
    \label{SIfig:Exp-setup}
\end{figure*}
California blackworms (\textit{Lumbriculus variegatus})  are  freshwater annelids that inhabit lake and river sediments (Fig.~\ref{SIfig:Exp-setup}c, blackworm in a channel). We obtained worms from Carolina Biological Supply and Eastern Aquatics and  maintained them in plastic  containers ($35 \times 20 \times 12$ cm) with  filtered, treated water and stored in a refrigerator at $ 4^\circ C$, following established protocols~\cite{martinez2021cuts}. Worms were fed tropical fish flakes once per week,  water was refreshed daily. And since blackworms are exempt from institutional animal care review, no additional approval was required. Before  experiments, worms were habituated for at least one hour in  room-temperature water ($\sim 21^\circ C$) under ambient lighting. 

%(length: $30.2\pm 7.4 mm$, diameter $D$:$0.6 \pm 0.1$ mm, mass: $7.0 \pm 2.4$ mg)
The quasi-2D experimental setup (Fig.~\ref{SIfig:Exp-setup}a) was fabricated from  3 mm thick acrylic sheets, laser-cut (using  Trotec  Speedy 100 laser cutter) and then bonded together to form two circular chambers connected by a straight channel. Both channel and chambers are open at the top. We  fixed the channel length  at $L_c=12 ~\textrm{cm}$, and varied the channel width ${W}$, as given in Table~\ref{tab:T1}.  For all our experiments, we selected worms with a body length of $L=5\pm 1 $ cm and diameter $D=0.5 \pm 0.1$ mm. The persistence length ($\ell_p$) of the worm, obtained from tangent-tangent correlations, is $6.5\,\mathrm{mm}$~\cite{sweeper-worm-collection-25}.  

Channels were filled with  filtered, treated water and a single worm was introduced.  Experiments were recorded at 30 fps for 1-2 hours using a top-mounted Logitech C920x (Logitech, Taiwan, China) webcam
 in a light box, without any external perturbations at room temperature (Fig.~\ref{SIfig:Exp-setup}b).
 
\begin{table}[h]
\centering
\begin{tabular}{|c|c|}
\hline
$W$ (mm) & $\widetilde{W}$ \\ \hline
1   & 2   \\ \hline
2  & 4  \\ \hline
3    & 6   \\
\hline
4   & 8   \\
\hline
6   & 12   \\
\hline
8   & 16   \\
\hline
\end{tabular}
\caption{Channel width $W$ and corresponding  scaled width $\widetilde{W}$ used in  worm experiments.}
\label{tab:T1}
\end{table}

\section{Data Analysis}
We quantified three observables from videos: the translocation (escape) time of the filament, the  center of mass of the filament, and the end-to-end distance.  Filaments were detected using standard thresholding and contour detection algorithms from the Python OpenCV library
(https://github.com/opencv/opencv-python). The raw images were first converted to grayscale and smoothed to reduce noise, after which a binary mask of the filament was obtained through adaptive thresholding. Contours of the segmented filament were then extracted frame-by-frame. These contours were further processed using a skeletonization algorithm from the Python scikit-image library
(https://github.com/scikit-image/scikit-image) to compute geometric properties. Although the worm exhibits 3D motion, the setup constrains vertical excursions; we therefore analyze  $\{x, y\}$ coordinates only. This approximation does  not  affect the escape time or center-of-mass dynamics because the out-of-plane fluctuations are suppressed.
%We used standard thresholding and contour detection algorithms from the Python OpenCV library (https://github.com/opencv/opencv-python) to detect the filament. The resulting contour was then processed using a skeletonization algorithm from the Python scikit-image library  to compute the properties.

%{\color{blue}Paulami Experimental details\\
%setup schematic}

\section{Simulation details}
\begin{figure*}[!h]
    \centering
\includegraphics[width=\textwidth]{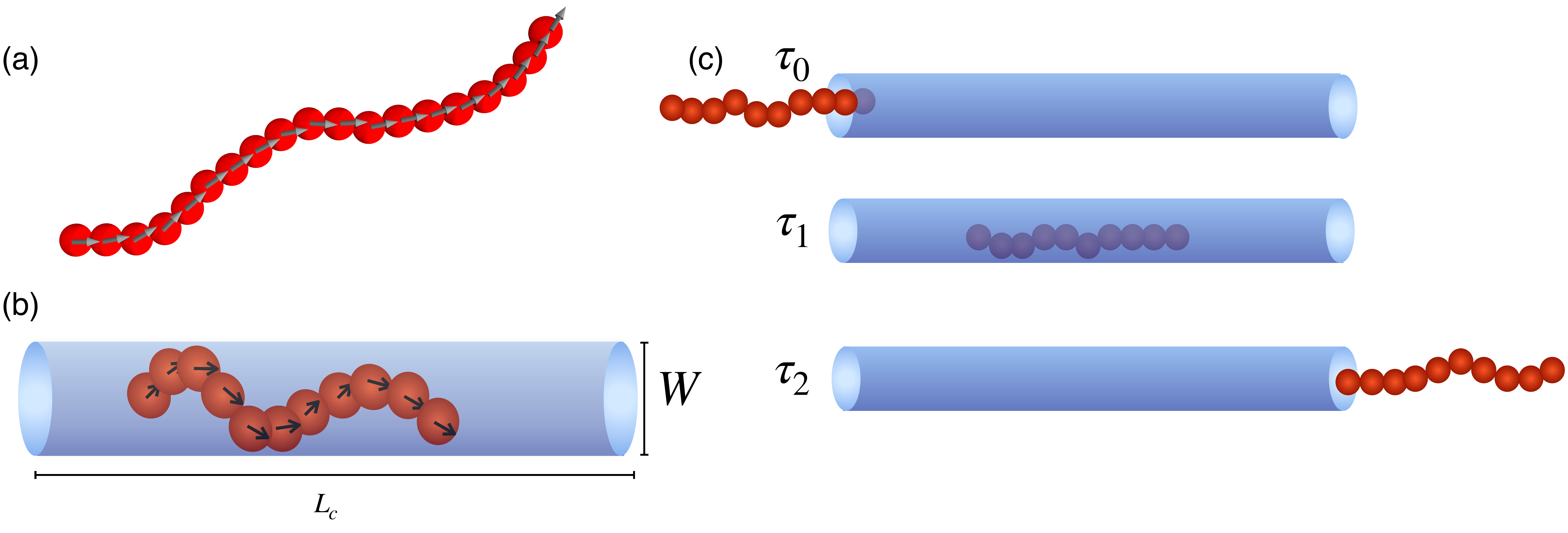}
    \caption{ \textbf{Active Polymer (AP) model.}  (a) Tangentially-propelled filament: each monomer propels with an active force ${f_p}$  directed  along the local tangent. (b) The active filament moves within an open-ended cylindrical channel of width $W$ and length $L_c$.  (c) Escape time measurement: $t_{esc}$  is measured  from $\tau_0$ (head entry, first monomer) to $\tau_2$ (tail exit, last monomer), including the residence interval $\tau_1$ during which the filament undergoes propulsion-driven motion, reorientations, and conformational fluctuations.%when the tip (first monomer) of the worm (polymer) enters the channel. The organism or polymer then remains within the channel for a time $\tau_1$, during which it propels forward and undergoes re-orientation and conformational fluctuations. Finally, the last monomer (tail end of the worm)  exits from the other side of the channel  at $\tau_2$.
    }
    \label{SIfig:Ttime}
\end{figure*}

{\it Active Polymer (AP) model.} We consider an  active polymer made of $N$ monomers with coordinates $r_i$, connected by harmonic springs with a bending potential,  confined within a cylindrical channel of length $L_c$ and diameter $W$ (Fig.~\ref{SIfig:Ttime}a)~\cite{prathyusha2018PRE}. 
The equation of motion for each monomer, $i$, is 
\begin{equation}
\gamma{{\dot {\bf r}}_i}=-\nabla_i({ U}^{WCA} + { U}^{Wall}+{U}^{stretch}{+ U}^{bend}
)+{\bf F}_i^{R}+ {\bf F}_i^{A} 
 \label{eq:eqofmotion}
\end{equation}
The steric interaction $ { U}^{WCA}$ between  all the monomers in the system is modeled via the Weeks-Chandler-Anderson potential \cite{weeks-chandler-71}, 
\begin{equation}
U^{WCA}(r)=4\varepsilon\left[\left(\frac{\sigma}{r}\right)^{12}-\left(\frac{\sigma}{r}\right)^{6}+\frac{1}{4}\right],
\label{eqn:wca}
\end{equation}
which vanishes  for  $r>2^{1/6}\sigma$. Here $\varepsilon$ measures the strength of the repulsive interaction, $\sigma$, the size of the particles,  and both are set to $1$. 
We  measure length in units of particle size $\sigma$, energies in $\varepsilon$, and units of time in $\tau=\sigma^2\gamma / \varepsilon$. 

%$\sigma$=1  is the size of the particles. 

We consider an open-ended channel of length $L_c$ and diameter $W$, Fig.\ref{SIfig:Ttime}(b). 
The confining wall interaction for each monomer is  given by, 
\begin{equation}
U^{Wall}(\lambda-|r|)=4\varepsilon\left[\left(\frac{\sigma}{\lambda-|r|}\right)^{12}-\left(\frac{\sigma}{\lambda-|r|}\right)^{6}+\frac{1}{4}\right],\label{eqn:wall}
\end{equation}
where $\lambda-|r|$ represents the separation distance between the monomer and the wall extending up to ($\lambda-|r|= 2^{\frac{1}{6}}\sigma$). After accounting for the finite size of the monomer, the available space for the monomer is defined as the effective width  $W=2(\lambda-\frac{\sigma}{2})$,  corresponding to the region where the wall repulsion becomes physically significant.

The bonded monomers of the polymer interact via a  harmonic potential, 
\begin{equation}
U^{stretch}(r)=\frac{k_b}{2}\big(r- R_0\big)^2 \label{eq:bond}.
\end{equation}
  Here $R_0=1.0\sigma$ is the equilibrium bond length, and $k_b=4000~k_BT/\sigma^2$
is the bond stiffness. These parameters make the chain effectively inextensible, diameter of the polymer $D= 1\sigma$ and polymer length $ L =N\sigma$.

Any connected triplet in the polymer experiences a bending interaction, which is given by the potential,
\begin{equation}
U^{bend}= \frac{\kappa }{2}\big(\theta -\theta_0 \big)^2
\end{equation}
 Here $\kappa$ is the bending stiffness and  $\theta$ is the angle between any consecutive bond vectors and $\theta_0$ is set to $\pi$. We vary $\kappa$ to change to flexibility  of the polymer.  \\
 
 ${\bf F}_i^{R}$ represents the stochastic force with zero mean and variance $2\gamma k_BT$ and we kept $k_BT=0.1 \varepsilon$. \\

In order to mimic the overall behavior of the worm we use a tangential active polymer model, where each bead is self-propelled along the local tangent of the polymer backbone~\cite{prathyusha2018PRE,prathyusha_softmatter-22,riseleholder-15}.   The last term represents self-propulsion force for the polymer, $ \mbox
{\bf F}_i^{A}=\frac{f_p}{2}({\bf \hat  t}_{i-1,i}+{\bf \hat  t}_{i,i+1}) $
Here, $f_p$ is the strength of the force and ${\bf \hat  t}_{i,i +1} = {\bf r}_{i,i +1} / {\bf r}_{i,i +1}$ is the unit tangent vector along the bond connecting beads i and i + 1. This is experienced by  the beads ($i=2,3,4... N-1$) of the polymer. 
The active forces for the end beads ($i=1$ and $i=N$), as they  have only  one nearest neighbor,   are \begin{equation}
\mbox {\bf F}_{1}^{A}=   \ \hat{\bf t}_{1}   f_{p} \; \& \; \;
 %\\
 {\bf F}_{N}^{A}=   \hat{\bf t}_{N-1} f_{p}
\label{activeforce-end}
\end{equation}

We set the number of monomers to $N = 30$  and the channel length to $L_c = 60$.  We vary the dimensionless diameter ${\widetilde{W}}$ in the range $[1.245 - 7]$ and the dimensionless persistence length $\ell_p/L$ in the range $[0.067- 1.5]$, which control geometric confinement and filament flexibility, respectively. The equations of motion \ref{eq:eqofmotion}
were integrated with time step, $\delta t=0.0001 \tau$.
\section{Translocation time $t_{\mathrm esc}$ analysis.}
%There are multiple ways to measure the translocation time, and in our manuscript, we employ the following rule. 
We define  a  successful translocation as an event in which a filament  enters the channel from one end and fully exits the opposite end within the observation window. The  translocation time or escape time $t_{\mathrm esc}$ is measured from  head entry of the filament to tail exit, i.e., $t_{\mathrm esc}=\tau_2-\tau_0$ (Fig.\ref{SIfig:Ttime}c), naturally capturing the time for the entire contour to traverse the channel. This definition follows established practice for driven translocation~\cite{yong2012driven}.

%the measurement of escape time starts at $\tau_0$, when the tip of the filament (or first monomer of the polymer) enters the channel. The filament resides within the channel for a duration $\tau_1$ and it interacts with the channel walls. At $\tau_2$ the tail-end (last  monomer of the polymer) escapes from the other end of the channel Fig.\ref{SIfig:Ttime}(c). This method naturally captures the time required for the entire contour length of the polymer to pass through the channel~\cite{yong2012driven}.

\section{Time evolution of the MSD exponent ($\beta(t)$) }
\begin{figure*}[!h]
    \centering
\includegraphics[width=\textwidth]{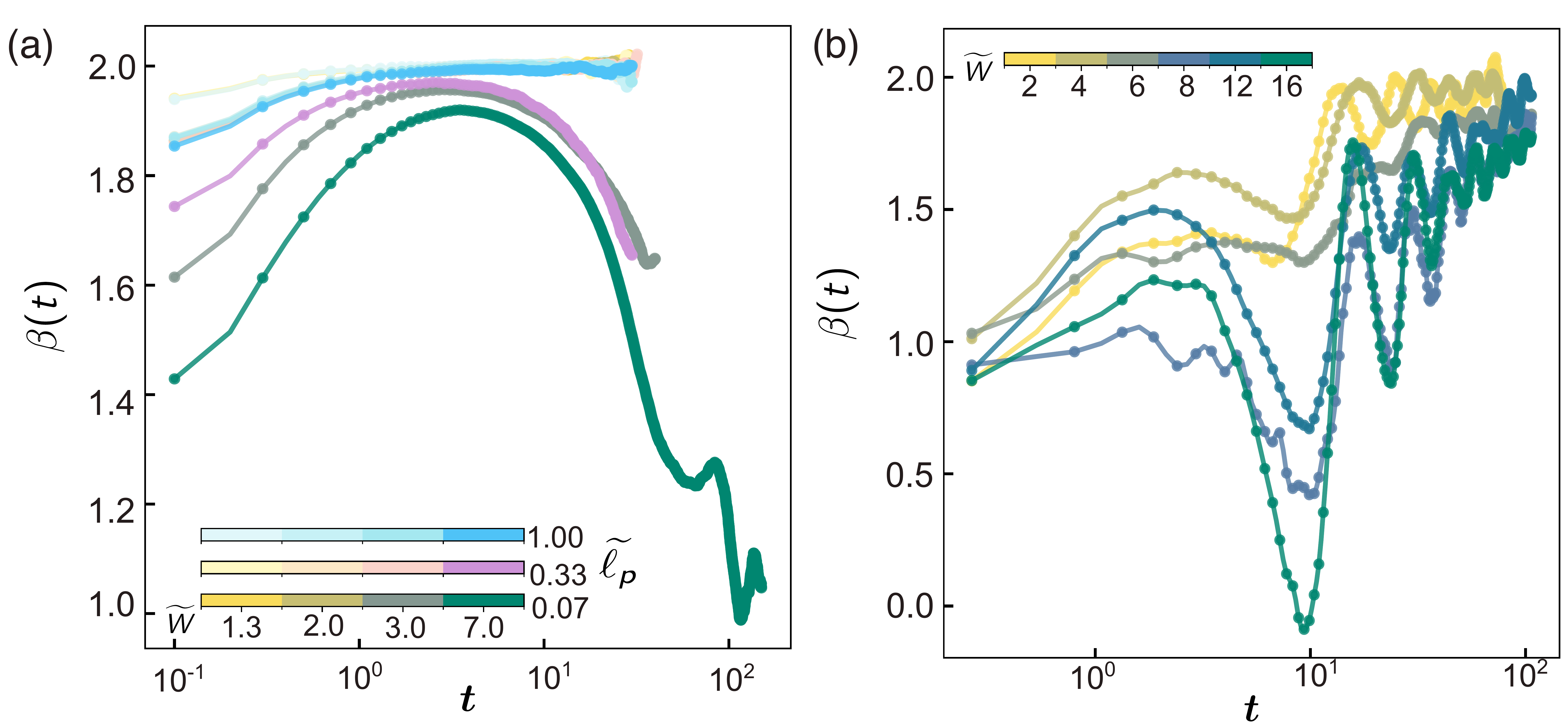}
    \caption{\textbf{ Local MSD exponent $\beta(t)$.}  (a) From AP simulations for different values of $\widetilde {\ell_p}$ and $\widetilde{W}$ (b) for worms in different confinement.}
    \label{beta-msd}
\end{figure*}

To characterize the transient dynamics of the filament in confinement, we compute the time-dependent MSD exponent $\beta(t)=\frac{\partial{ \textrm{ln(MSD}(t)}) }{\partial {\textrm{ln}(t)}}$, as shown in Fig.
\ref{beta-msd}.
Under strong confinement or for stiff polymers, $\beta \sim 1.9$, throughout the translocation, consistent with nearly ballistic, directed motion (Fig.
\ref{beta-msd}a). In contrast, when $\widetilde{W} > 2 $ and  polymers are flexible, $\beta(t)$ becomes non-monotonic: it rises to $\gtrsim1.8$ at intermediate times, then crosses over to {$\beta \sim 1.1 $ } as reorientations decorrelate directional persistence. 

For worms, $\beta$ follows an overall rising trend, exhibiting 
a long-term superdiffusive behavior ($\beta \sim 1.8$) consistent with simulations, with a short term intermediate  subdiffusive plateau $\beta \in[0,1.5]$ arising from intermittent pauses/reorientations; this plateau is absent in the AP model due to constant propulsion (~Fig.
\ref{beta-msd}b).

%for the worm indicating persistent, directed locomotion inside the channel. At intermediate regime, $\beta$ value vary between [0-1.5] suggesting that there is a plateau in MSD. This can be attributed to periods of dwelling, during which the worm exhibits intermittent pauses, reorientation maneuvers, or transient substrate contacts that prevent it from moving forward. Although the active-polymer (AP) model reproduces the long-time persistent (superdiffusive) motion
% it does not reproduce the subdiffusive plateaus observed experimentally. The constant tangential propulsion in AP continuously drives the polymer and thereby suppresses transient dwelling observed for worms. We leave a detailed investigation of these dwelling behaviour to future work.

\section{Saturation time of radial MSD}
\begin{figure*}[!h]
    \centering
\includegraphics[width=0.5\textwidth]{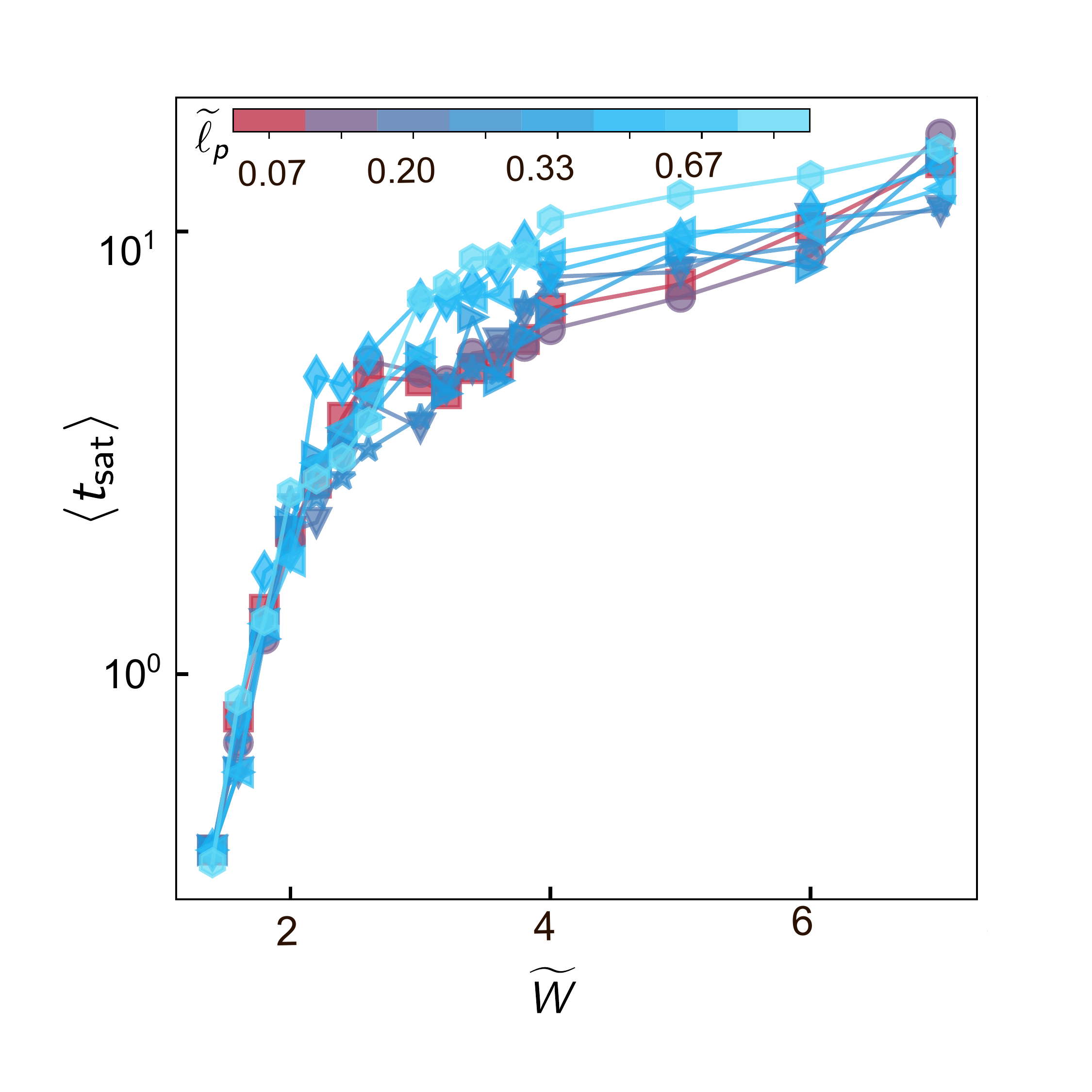}
    \caption{ The saturation time ($t_{\textrm {sat}}$)  of rMSD as a function of channel width; initially increases with $\widetilde{W}$  and then saturates, showing no dependence on filament stiffness.}
    \label{SIfig:TSAT}
\end{figure*}

We  quantify transverse exploration by the radial mean-squared displacement
  $\textrm{rMSD}
= \langle \frac{1}{N} \sum_{i=1}^{N} 
| \mathbf{r}_i^{\perp}(t+\tau) - \mathbf{r}_i^{\perp}(\tau) |^2 \rangle$, with $\mathbf{r}_i^{\perp}(t) = \{ y_i(t), z_i(t) \}$ (main text Fig.2b). The rMSD grows and then saturates at  $\textrm{rMSD}_{\mathrm {sat}}$. %$\Delta {\mathbf r}^2_{\perp, {\mathrm sat}}$ grows systematically with channel width due to activity-driven exploration of the confinement boundary and remains largely independent of filament flexibility. 
 $\textrm{rMSD}_{\mathrm {sat}}$ increases with channel width, indicating enhanced activity-driven radial exploration of monomers and  is largely independent of filament stiffness.  We find that $\textrm{rMSD}\sim \widetilde{W}^{2.5}$, steeper than the $W^2$ scaling expected for a passive tracer. The super-quadratic growth reflects persistent self-propulsion and wall-following (boundary dwelling and sliding), which increase the fraction of the contour near the walls and produce larger cross-sectional excursions than geometry alone would predict. 
%transverse displacements scale faster than the geometric expectation under passive confinement.
 The time to reach its saturation value, referred to as the saturation time $t_{\mathrm {sat}}$ also increases with $\widetilde{W}$ and then saturates beyond $\widetilde{W} > 3.0$, Fig.~\ref{SIfig:TSAT} indicating confinement-limited dynamics. Notably, $t_{\mathrm{sat}}$ shows no dependence on stiffness (data collapse across $\widetilde{\ell_p}$), indicating geometry-confinement and self-propulsion, not bending fluctuations, set the timescale. 
Although $t_{\mathrm{sat}}<< t_{\mathrm{th}}$, it becomes relevant in the stiff limit. 

%Unlike passive polymers, where conformational freedom slows transport, active filaments maintain axial motion via propulsion, though flexibility can still cause delays.

\section {End to End vector Analysis and Reorientation dynamics}
\begin{figure*}[!h]
    \centering
\includegraphics[width=\textwidth]{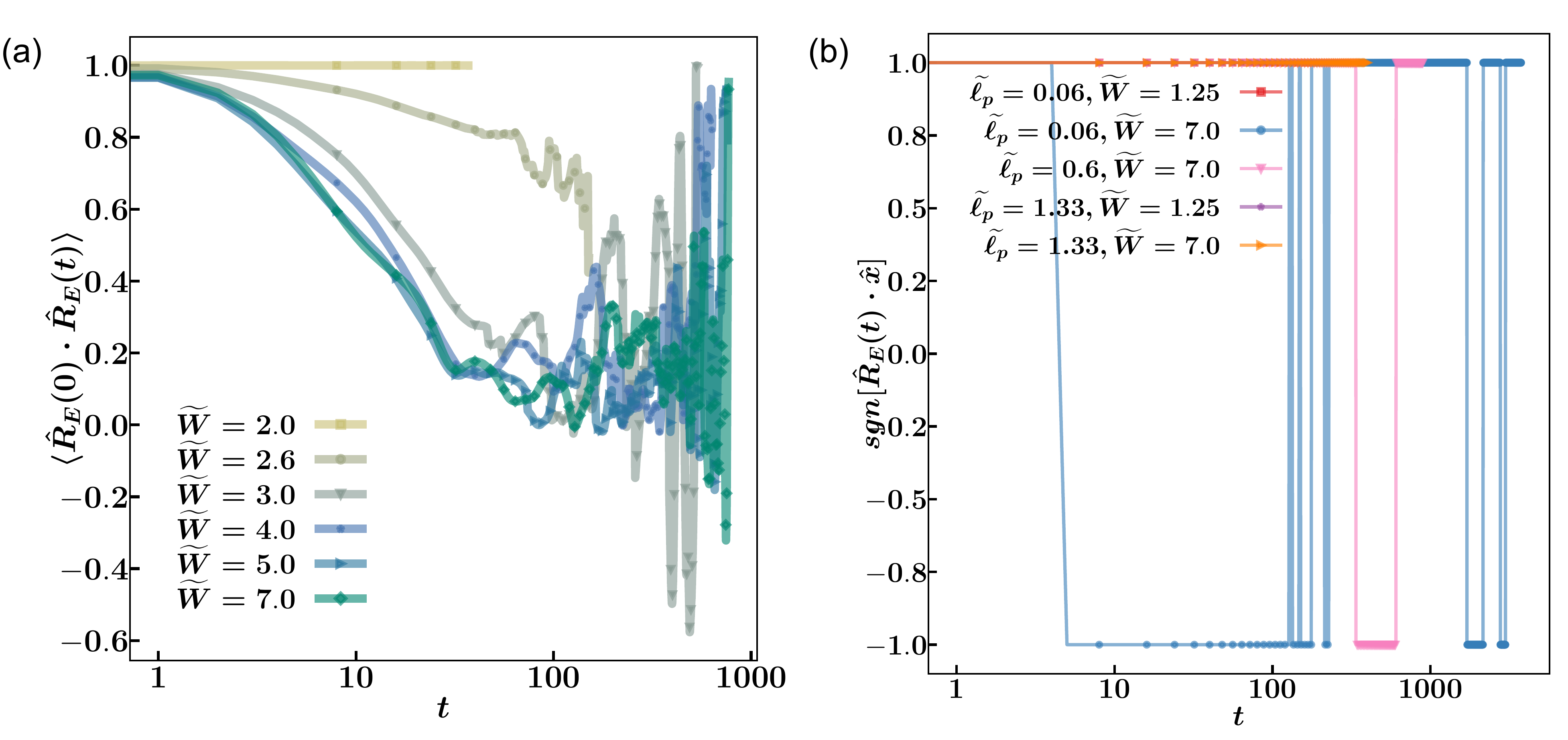}
    \caption{\textbf{ Orientation dynamics in confinement.}
   (a) Autocorrelation of end-to-end vector ${\bf R}_E(t)\cdot{\bf R}_E(0)$:   near-constant under strong confinement ($\widetilde{W} \leq 2.0$); and compressed-exponential-like decay in wider channel due to re-orientations. (b) Axial alignment signal ${sgn}[{\bf \hat R}_E(t) \cdot \hat{\bf x}]$: step-like toggles mark reorientation events used to compute reorientation counts $N_{RE}$.}
    \label{SIfig:RE-corr}
\end{figure*}

We analyze the end-to-end vector ${\bf R}_{E}(t)$=${\bf R}_N(t)-{\bf R}_1(t)$ to probe orientation, where   ${\bf R}_1(t)$, ${\bf R}_N(t)$ are co-ordinates of the  first and last monomers of the polymers respectively. The autocorrelation of the end-to-end vector for active filaments under various confinements is shown in Fig~\ref{SIfig:RE-corr}(a), computed over the interval when the entire polymer is inside the channel. In strong confinement ($\widetilde{W} \leq 2.0$), the end-to-end vector remains close to 1, and the correlation shows negligible decay over time. When $\widetilde{W}>2.0$, the correlation decays in a compressed exponential manner, reflecting abrupt reorientations due to persistent propulsion and confinement-induced re-alignment.

 %To quantify these reorientations of the end-to-end vector, we first calculated the time evolution of the end-to-end vector of the polymer, ${\bf R}_{E}(t)$=${\bf R}_N(t)-{\bf R}_1(t)$, where   ${\bf R}_1(t)$, ${\bf R}_N(t)$ are co-ordinates of the  first and last monomers of the polymers respectively. 
  %To quantify these reorientations of the polymer, we end-to-end vector
 %The 
 %correlation of ${\bf R}_{E}(t)$ for different confinement widths is shown in figure  \ref{SIfig:RE-corr}a.  Figure \ref{SIfig:RE-corr}a shows the auto-correlation of ${\bf R}_{E}(t)$  
%for different confinement widths, computed during the interval when the entire polymer is inside the channel.For $\widetilde{W} < 2$, during the entire period, the 
 
 To quantify the reorientation of the polymer within the confining channel, we measure the alignment  of ${\bf R}_{E}(t)$, with the channel axis  $\hat{\bf x}$, i.e., ${sgn}[{\bf \hat R}_E(t) \cdot \hat{\bf x}]$, Fig~\ref{SIfig:RE-corr}(b). Each sign change indicates a reorientation of $\bf{R}_E$ relative to $\hat{\mathbf{x}}$, and the total number of such sign changes over the translocation trajectory provides a direct number of the  axial reversals of the filament (main text Fig 2(c)). 

%We have not included fluid flows in our simple model for an enzyme cluster.
%We discuss here, how the effect of fluid flows can be incorporated. the diffusiophoretic interactions contribute stresses equal to $\rho \mathbf{E}$.  

\newpage
\section {SI Movies}
\noindent\textbf{Notes.} Simulation movies were rendered from frames recorded every $0.1\,\tau$; playback is accelerated $8\times$ for visualization.
\begin{enumerate}
  \item {\bf SM1.} Comparative translocation of blackworms (\textit{Lumbriculus variegatus}) across confinements ($\widetilde W=2,4,12$). Strong confinement yields faster escape.  
  \item {\bf SM2.} Failed translocation (worm). A worm in a wide channel ($\widetilde W=16$) reorients and exits from the entry side, illustrating a failed event.
 
  \item {\bf SM3.} Successful translocation with reorientation (worm). A worm in a wide channel ($\widetilde W=16$) undergoes body reorientation during a successful escape.%Successful translocation and reorientation of a worm:~Experimental movie of a worm in a confined channel~($\widetilde{W}=16$)  undergoing body reorientation during a successful translocation. 
   \item{\bf SM4.} Strong confinement (active polymer). Flexible polymer with $\widetilde{\ell}_p=0.067$ at $\widetilde W=1.245$. All beads propel along the local tangent; no axial reorientations are observed.
%The movie shows the flexible active filament in strong confinement :  $\widetilde{\ell}_p=0.067 $, ${\widetilde W}= 1.245$. All the beads translocate with the force and direction and no re-orientation is observed. (In all  simulation movies: Frames were recorded every 0.1 $\tau$ for the movie. For visualization, the playback speed was accelerated by a factor of 8.)
\item{\bf SM5.} Strong confinement (active polymer). Stiffer polymer with $\widetilde{\ell}_p=1.33$ at $\widetilde W=1.245$. Dynamics closely match the flexible case, indicating weak stiffness dependence in this regime.%The video shows a stiffer active filament in strong confinement: $\widetilde{\ell}_p=1.33 $, ${\widetilde W}= 1.245$. The dynamics of the  stiffer filament closely resemble those of a flexible filament in strong confinement, indicating that flexibility has little effect on translocation behavior in this regime.

\item {\bf SM6.} Wide channel, de Gennes–like regime (active polymer). Flexible polymer with $\widetilde{\ell}_p=0.067$ at $\widetilde W=7.0$ exhibits frequent reorientations and shape changes.%The video demonstrating the re-orientation of  flexible active filament in wider channel (de Gennes-like): $\widetilde{\ell}_p=0.067 $, ${\widetilde W}= 7$. 
      
  \item {\bf SM7.} Stiffer polymer with $\widetilde{\ell}_p=1.33$ at $\widetilde W=7.0$ maintains an extended, axis‑aligned, Odijk-like conformation with short wall deflections.%The video explains the translocation  of  stiffer active filament in wider channel exhibiting an extended conformation (Odijk-like): $\widetilde{\ell}_p=1.33 $, ${\widetilde W}= 7$.

\item {\bf SM8.} Failed translocation (active polymer). Flexible polymer with $\widetilde{\ell}_p=0.067$ at $\widetilde W=3.0$ repeatedly reorients within the channel and fails to exit the far end.%Movie showing a failed translocation event due to reorientation of the polymer within the channel:~$\widetilde{\ell}_p=0.067 $, ${\widetilde W}= 3$.
  
\end{enumerate}

%

%\bibliography{active-polymer}
\end{document}